\documentstyle[preprint,eqsecnum,aps]{revtex}

\begin{document}
\title{Small Effects in Astrophysical Fusion Reactions}
\author{A.B. Balantekin$^a$, C. A. Bertulani$^b$ and M.S. Hussein$^c$}
\address{$^a$ University of Wisconsin, Department of Physics\\
Madison, WI 53706, USA. E-mail: baha@nucth.physics.wisc.edu\\
$^b$ Instituto de F\'\i sica, Universidade Federal 
do Rio de Janeiro \\
21945-970 \ Rio de Janeiro, RJ, Brazil. E-mail: 
bertu@if.ufrj.br\\
$^c$ Universidade de S\~ao Paulo, Instituto de F\'\i sica\\
05389-970 \ S\~ao Paulo, SP, Brazil. E-mail: hussein@if.usp.br}
\maketitle

\begin{abstract}
We study the combined effects of vacuum polarization, relativity,
Bremsstrahlung, and atomic polarization in nuclear reactions of
astrophysical interest. It is shown that these effects do not solve the
longstanding differences between the experimental data of astrophysical
nuclear reactions at very low energies and the theoretical calculations
which aim to include electron screening.
\end{abstract}


\section{Introduction}

Understanding the dynamics of fusion reactions at very low energies is
essential to understand the nature of stellar nucleosynthesis. These
reactions are measured at laboratory energies and are then extrapolated to
thermal energies. This extrapolation is usually done by introducing the
astrophysical S-factor: 
\begin{equation}
\sigma \;(E)={\frac 1E}\;S(E)\;\exp \left[ -2\pi \eta (E)\right] \ ,
\label{sig1}
\end{equation}
where the Sommerfeld parameter, $\eta (E),$ is given by$\;\eta
(E)=Z_1Z_2e^2/\hbar v.$ Here $Z_{1,}$ $Z_2,$ and $v$ are the electric
charges and the relative velocity of the target and projectile combination,
respectively.

The term $\exp (-2\pi \eta )\ $is introduced to separate the exponential
fall-off of the cross-section due to the Coulomb interaction from the
contributions of the nuclear force. The latter is represented by the
astrophysical S-factor which is expected to have a very weak energy
dependence. The form given in Eq. (2.1) assumes that the electric charges on
nuclei are ``bare''. However, neither at very low laboratory energies, nor
in stellar environments this is the case. In stars the bare Coulomb
interaction between the nuclei is screened by the electrons in the plasma
surrounding them. A simple analytic treatment of plasma screening was
originally given by Salpeter\ \cite{Sal54}. In most cases of astrophysical
interest Salpeter's treatment still remains to be a sufficient approximation 
\cite{Bah96}. In the very low energy laboratory experiments the bound
electrons in the projectile or the target may also screen the Coulomb
potential as the outer turning point gets very large (%
\mbox{$>$}
500 fm). As experimental techniques improve one can measure the cross
section in increasingly lower energies where the screened Coulomb potential
can be significantly less than the bare one. This deviation from the bare
Coulomb potential should manifest itself as an increase in the astrophysical
S-factor extracted at the lowest energies. This enhancement was indeed
experimentally observed for a large number of systems \cite
{Eng88,Eng92,Ang93,Pre94,Gre95}. The screening effects of the atomic
electrons can be calculated \cite{Ass87} in the adiabatic approximation at
the lowest energies and in the sudden approximation at higher energies with
a smooth transition in between \cite{Sho93}.

In the adiabatic approximation one assumes that the velocities of the
electrons in the target are much larger than the relative motion between the
projectile and the target nuclei. In this case, the electronic cloud adjusts
to the ground-state of a ``molecule'' consisting of two nuclei separated by
a time-dependent distance $R(t)$, at each time instant $t.$ Since the
closest approach distance between the nuclei is much smaller than typical
atomic cloud sizes, the binding energy of the electrons will be given by the
ground-state energy of the $Z_P+Z_T$ atom, $B(Z_P+Z_T)$. Energy conservation
implies that the relative energy between the nuclei increases by $%
U_e=B(Z_T)-B(Z_P+Z_T)$. This energy increment increases the fusion
probability. In other words, the fusion cross section measured at laboratory
energy $E$ represents in fact a fusion cross section at energy $E+U_e$, with 
$U_e$ being called by the {\it screening potential. }Using eq. (1.1), one
gets

\begin{equation}
\sigma \;(E+U_e)\cong \exp \left[ \pi \eta (E)\frac{U_e}E\right] \ \sigma
(E)\;,  \label{sig1}
\end{equation}
where one assumes that the factor $S(E)/E$ varies much slower with $E$, as
compared to the energy dependence of $\exp \left[ -2\pi \eta (E)\right] $ .

The exponential factor on the right-hand-side of eq. (1.2) is the
enhancement factor due to the screening by the atomic electrons in the
target. For light systems the velocity of the atomic electrons is comparable
to the relative motion between the nuclei. Thus, a dynamical calculation has
to be done for the effect of atomic screening \cite{Sho93}. However, the
screening potential $U_e$ obtained from a dynamical calculation cannot
exceed that obtained in the adiabatic approximation because the dynamical
calculation includes atomic excitations which reduce the energy transferred
from the electronic binding to the relative motion.

Contributions from the nuclear recoil caused by the atomic electrons are
expected to further increase the screening effect for asymmetric systems 
\cite{Sho93,Bri81} . In almost all the cases observed screening effects are
found to be equal to or more than the theoretical predictions. Recently
including improved energy loss data for atomic targets is shown to lead
agreement between theory and data \cite{Ban96,newkarl}, however the
situation is still not resolved for molecular and solid targets. Electron
screening enhancement was not observed for the heavier symmetric system $%
^3He(^3He,2p)^4He$ \cite{Kra87} which is expected to have about 20\%
enhancement at the energies studied. Recent measurements \cite{rol} have not
yet clarified the effects of electron screening in this reaction. A
mechanism which reduces the screening enhancement for this system (and
possibly for other systems with large values of $Z_1Z_2$ and the reduced
mass) seems to be needed.

In this article we show that the contributions from the polarization of the
vacuum, relativity, Bremsstrahlung, and atomic polarization cannot achieve
this task. The motivation for this investigation is that $U_e/E,$ appears
multiplied by a large number, $\eta (E),$ in the exponential factor of eq.
(1.2). One needs only a small value of $U_e$ to obtain a sizeable
enhancement factor; typically $U_e/E\sim 0.001$ . The effects of the vacuum
polarizability was previously investigated in Ref. \cite{Bau77} for elastic
scattering below the Coulomb barrier and in Ref. \cite{son} for subbarrier
fusion reactions using the formalism developed by Uehling \cite{Ueh35}.
Effects of vacuum polarization in $^{12}C-^{12}C\;$scattering at 4 MeV was
subsequently experimentally observed \cite{Vet89}. Other small effects in
elastic scattering at low energies have also been studied by several authors 
\cite{Hus84,Hus90,Ber94}. They have also been studied in the context of
astrophysical reactions in refs. \cite{Kam94,DeF94}. However, to our
knowledge, the other effects have not been studied.

In section 2 we study the effects of vacuum polarization, relativity,
Bremsstrahlung, and atomic polarization for the astrophysical reactions
listed in ref. \cite{rol}, and for which a set of ``experimental '' values
of screening energies $\Delta U_e$ are given. These experimental values are
chosen so that the eq. (1.1) reproduces the enhancement of the fusion cross
sections at very low energies. In section 3 we present our conclusions.

\section{Small effects in thermonuclear reactions {\it \ }}

To calculate the fusion cross section corrections we use for simplicity the
(s-wave) WKB penetrability factor

\begin{equation}
P(E)={\exp }\left[ {-}\frac 2\hbar \int_{R_n}^{R_C}dr\ |p(r)|\right] \ ,
\label{tmat}
\end{equation}
where $p(r)$ is the (imaginary) particle momentum inside the repulsive
barrier. The corrected fusion cross section is given by

\begin{equation}
\sigma =\ \sigma _C\;.\;R\;,  \label{tmat}
\end{equation}
where $\sigma _C$ is the pure Coulomb repulsion cross section, and $%
R=P_{C+\alpha }(E)/P(E),$ with $\alpha =\{$scr, VPol, rel, Brems, At\}, are
the corrections due to atomic screening, vacuum polarization, relativity,
Bremsstrahlung, and atomic polarization, respectively.

The atomic screening effect is calculated using $|p(r)|=\sqrt{2m\left[
V_c(r)-E-U_e\right] }$, where $E$ is the relative energy between the nuclei.
The atomic screening potential, assumed to be a constant function of $r$
(valid for $r\ll a_B=0.529\;\AA $), is given by $U_e$.

\subsection{Vacuum Polarization}

Vacuum polarization {\bf increases} the electromagnetic potential between
two like charges. Like the Coulomb potential itself, the increase due to
vacuum polarization is also proportional to the product of the charges \cite
{Ueh35}. Vacuum polarization contribution increases almost exponentially as
the two charges get closer. The Coulomb interaction is smaller for
asymmetric systems than for symmetric systems of comparable size. On the
other hand, the nuclear force tends to extend farther out for asymmetric
systems because of the extra neutrons. Consequently for asymmetric systems
the very tail of the nuclear force can turn the relatively weak Coulomb
potential around to form a barrier at a considerable distance from the
nuclear touching radius. For symmetric systems, however, the location of the
barrier is further inside where the vacuum polarization contribution is
stronger. We show that the resulting increase in vacuum polarization is
nevertheless not sufficiently large to make an appreciable contribution to
the extracted astrophysical S-factor. For light symmetric systems with small
values of $Z_1Z_2$ this effect should be negligible. Indeed, for the $pp$
reaction the vacuum polarization contribution was shown to be very small 
\cite{Kam94}. Similarly the measured S-factor for the $d(d,p)^3H$ reaction 
\cite{Gre95} agrees well with theoretical calculations of atomic screening 
\cite{Bra90}. On the other hand one may expect that already for the $%
^3He(^3He,2p)^4He$ reaction the increase in the potential due to the vacuum
polarization could be large enough to counter the decrease due to electron
screening. We show that this is not the case.

The vacuum polarization potential is according to Uehling \cite{Ueh35} given
by

\begin{equation}
V_{Pol}(r)=\frac{Z_1Z_2e^2}r\;\frac{2\alpha }{3\pi }\;I%
{2r \overwithdelims() \lambda _e}
\;,
\end{equation}
where $\alpha =1/137$ is the fine structure constant, and $\lambda _e=386$
fm is the Compton wavelength of the electron. The function $I(x)$ is given by

\begin{equation}
I(x)=\int_1^\infty e^{-xt}\;(1+\frac 1{2t^2})\;\frac{\sqrt{t^2-1}}{t^2}%
\;dt\;.
\end{equation}
As shown by Pauli and Rose \cite{Pau36} this integral can be rewritten as

\begin{equation}
I(x)=\alpha (x)K_0(x)+\beta (x)K_1(x)+\gamma (x)\int_x^\infty K_0(t)\;dt\;,
\end{equation}
where

\begin{eqnarray}
\alpha (x) &=&1+\frac 1{12}x^2\;,\;\;\;\;\;\;\;\;\;\;\;\;\;\;\beta (x)=-%
\frac 56x(1+\frac 1{10}x^2)\;,  \nonumber \\
\gamma (x) &=&\frac 34x(1+\frac 19x^2)\;,\;\;\;\;\;\;\;\;\;\;\;\text{with}%
\;\;x=2r/\lambda _e\;.
\end{eqnarray}
In Ref. \cite{Bau77} it was shown that the modified Bessel functions $K_0$
and $K_1$ as well the integral over $K_0$ can be expanded in a very useful
series in Chebyshev polynomials which converge rapidly and for practical
purposes only a few terms $(\approx 5-10)$ is needed, allowing a very fast
and accurate computation of the Uehling potential.

In Figure 1 we plot the Coulomb potential and the vacuum polarization
potential for $Z_1Z_2=1$. Both the Coulomb potential and the screening
potential scale with the product $Z_1Z_2$. However, the vacuum polarization
potential has a stronger dependence on the nuclear separation distance.

The limits of the integral are the nuclear radius, $R_n,$ where the nuclear
fusion reaction occurs, and the classical turning point in the Coulomb
potential, $R_C=Z_1Z_2e^2/E^{\prime },$ where $E^{\prime }=E+U_e.$ At very
low energies the inferior limit $R_n$ is not important when vacuum
polarization is neglected (the exponential factor in Eq. (1.2) can be
obtained with $V_{pol}=0,$\ and $R_n\rightarrow 0,$ in Eq. (2.1)). However,
since the vacuum polarization potential has a strong dependence on the
nuclear separation distance, being much stronger at shorter distances, its
effect is very much dependent on the choice of this parameter. For all
reactions with the deuteron we use the ``deuteron radius'', $R_n=4.3$ fm$,$
corresponding to an average distance value associated with matrix elements
involving the deuteron. For the other reactions we use $R_n$ values given in
the third column of Table I. In the 4th row we show the ratio between the
penetrability factor through the Coulomb barrier and the penetrability
factor including atomic screening, $P_{C+Scr}(E)/P_C(E)$. In the 5th row we
show the effect of vacuum polarization, $P_{C+VPol}(E)/P_C(E)$. The energy $%
E $ chosen is the lowest experimental energy for each reaction. The atomic
screening corrections $U_e$ were calculated in the adiabatic approximation,
given by the differences in electron binding energies between the separated
atoms and the compound atom \cite{Sho93}. We see that the effect of vacuum
polarization is small, but non-negligible for some reactions. Moreover, it
increases the discrepancy between the value of the screening potential
required to explain the experimental data and the theoretical calculations
of this potential as illustrated in Table I.

\subsection{Relativistic effects}

A classical Hamiltonian may be written which contains relativistic effects
to first order in $1/mc^2$. This Darwin Hamiltonian takes the following form
in the center of mass system

\begin{equation}
E=\frac{p^2}{2m_0}+\frac{Z_pZ_Te^2}r-\frac{p^4}{8c^2}\left( \frac 1{m_P^3}+%
\frac 1{m_T^3}\right) +\frac{Z_PZ_Te^2}{2m_Pm_Tc^2}\left( \frac{p^2+p_r^2}r%
\right) \;,
\end{equation}
where $m_0=m_Pm_T/(m_P+m_T)$ is the reduced mass, and $p_r$ is the radial
component of the relative momentum. In a head-on collision,

\begin{equation}
E=\frac{p^2}{2m_0}+\frac{Z_pZ_Te^2}r-\frac{p^4}{8c^2}\left( \frac 1{m_P^3}+%
\frac 1{m_T^3}\right) +\frac{Z_PZ_Te^2}{m_Pm_Tc^2}\frac{p^2}r\;.
\end{equation}

The solution of this equation yields, for $R_n\leq r\leq R_C,$

\begin{eqnarray}
|p| &=&\left( 2\beta \right) ^{-1/2}\;\left[ -\left( \alpha +\gamma \right) +%
\sqrt{\left( \alpha +\gamma \right) ^2+4\beta \left( V_C-E\right) }\right]
^{1/2},  \nonumber \\
\alpha &=&1/2m_0\;,\;\;\;\beta =\left( 1/m_P^3+1/m_T^3\right)
/8c^2\;,\;\;\;\gamma =Z_PZ_Te^2/\left( m_Pm_Tc^2r\right)
\end{eqnarray}

The correction due to relativity is given in the 6th column of Table I.
Although the correction in the momentum $|p|$ is of the order of $10^{-6}$,
the penetrability is enhanced by an amount of order of $10^{-3}$ as compared
to the penetrability with only the Coulomb interaction.

\subsection{Bremsstrahlung}

The energy emitted by Bremsstrahlung per frequency interval $d\omega $ and
solid angle element $d\Omega $ is

\begin{equation}
dE_{Br}(\omega )=d\omega d\Omega \frac{\omega ^2}{c^3}\left| \int_{-\infty
}^\infty \frac{dt}{2\pi }\;\text{e}^{i\omega t}\sum_jq_j\;\text{e}^{-ik%
\widehat{{\bf n}}.{\bf r}_j(t)}\left[ {\bf v}_j(t)\times \widehat{{\bf n}}%
\right] \right| ^2\;.
\end{equation}
where $k=\omega /c$ and the direction of observation $\widehat{{\bf n}}={\bf %
k/}k$. The sum goes over the charges $q_j$, positions ${\bf r}_j$, and
velocity ${\bf v}_j$ of the moving particles. In the long-wavelength
approximation

\begin{equation}
\;\text{e}^{-ik\widehat{{\bf n}}.{\bf r}_j(t)}{\bf \;}=1-(ik)\;\widehat{{\bf %
n}}.{\bf r}_j+...\;.
\end{equation}
and

\begin{equation}
dE_{Br}(\omega )=d\omega d\Omega \;A_R\;\frac{\omega ^2}{c^3}\left| {\bf d}%
(\omega )\times \widehat{{\bf n}}-\frac{ik}2{\bf Q}(\omega )\times \widehat{%
{\bf n}}+...\right| ^2\;.
\end{equation}
where, in the center of mass system, with relative positions and velocities
given by ${\bf r}(t)$ and ${\bf v}(t)$, respectively, and

\begin{eqnarray}
{\bf d}(t) &=&f_1e^2{\bf v}(t)\;,  \nonumber \\
{\bf Q}(t) &=&f_2e^2\left[ \widehat{{\bf n}}{\bf .r}(t)\right] {\bf v}(t)\;,
\nonumber \\
f_\lambda &=&A_R^{\lambda -1}\left( \frac{Z_P}{A_P^\lambda }-\frac{Z_T}{%
A_T^\lambda }\right) \;,  \nonumber \\
A_R &=&\frac{A_PA_T}{A_P+A_T}\;.
\end{eqnarray}

In a head-on collision

\begin{eqnarray}
r &=&a_0(\cosh \xi +1)\;,  \nonumber \\
t &=&\omega _0(\sinh \xi +\xi )\;,  \nonumber \\
a_0 &=&\frac{Z_PZ_Te^2}{2E}\;,\;\;\;\;\omega _0=\frac{a_o}{\text{v}}\;.\;\;\;
\end{eqnarray}
${\bf d}(\omega ),\;{\bf Q}(\omega )$ are the Fourier transforms of ${\bf d}%
(t)$ and ${\bf Q}(t)$, respectively, which can be calculated analytically.
The final result, after an integration over $\Omega $, is

\begin{equation}
\frac{dE_{Br}(\omega)}{d\omega}=\frac 4\pi \hbar \omega _0\alpha \left( 
\frac{{\rm v}}c\right) ^2A_R^2\;\left[ f_1^2h_1+f_2^2\;A_R^2\;\left( \frac{%
{\rm v}}c\right) ^2h_2\right] \;,
\end{equation}
where

\begin{eqnarray}
h_1 &=&\frac{2z}3\;\text{e}^{-\pi z}\;\left[ K_{iz}^{\prime }(z)\right]
^2,\;\;\;\;\;\;\;\;h_2=\frac{2z}{15}\;\text{e}^{-\pi z}\;\left[
K_{iz}(z)\right] ^2,  \nonumber \\
z &=&\omega /\omega _0\;,  \nonumber \\
K_{iz}(y) &=&\frac 12\int_0^\infty \text{e}^{-y\cosh \xi }\cos (z\xi )\;d\xi
\;,\;\;\;\;\;K_{iz}^{\prime }(y)=\frac{dK_{iz}(y)}{dy}\;.
\end{eqnarray}

The above results give the Bremsstrahlung due to the incoming branch of the
trajectory only, since we are interested in the energy loss until the fusion
occurs. The result for the full trajectory, including the outgoing branch is
obtained by replacing the lower limit of the integral in eq. (2.16) by $%
-\infty $.

The frequencies $z\ll 1$ $(\omega \ll \omega _0)$ dominate the spectrum and
we can replace the functions $K_{iz}(z)$ and $K_{iz}^{\prime }(z)$ by their
approximate values for low $z$. The integration of eq. (2.15) over $\omega$
is straightforward, and we get

\begin{equation}
E_{Br}=\frac 4{3\pi ^2}\hbar \omega _0\alpha \left( \frac{{\rm v}}c\right)
^2A_R^2\;\left\{ f_1^2+\frac 2{5\pi ^2}f_2^2A_R^2\left( \frac{{\rm v}}c%
\right) ^2\left[ \left( \frac 32-\ln (2\pi )\right) ^2+\zeta (2,2)\right]
\right\} \;,
\end{equation}
where $\zeta (2,2)=0.64493...$ is the Riemann's Zeta function for a
particular value of its argument.

The results of energy loss by Bremsstrahlung are given in the 7th row of
table 1, where $R_{Br}=P_{C+Br}/P_C$ is calculated by using eq. (1.2), with $%
U_e$ replaced by ($-E_{Br}$). It is larger for the systems with a large
effective dipole charge $f_1$, since the quadrupole radiation is smaller by
a factor $($v$/c)^2.\;$However, even for the systems with a larger $f_1,$
the Bremsstrahlung correction is of order of $10^{-3}$.

\subsection{Atomic polarizability}

The virtual excitations of the atomic electrons in the target yields in
second order perturbation theory a polarization potential given by

\begin{equation}
V_{at}=-\sum_{n\neq 0}\frac{\left| \left\langle 0\left| V_C({\bf r,R}%
)\right| n\right\rangle \right| ^2}{E_n-E_0}\;,
\end{equation}
where

\begin{eqnarray}
V_C({\bf r,R}) &=&\sum_i\frac{Z_Pe^2}{\left| {\bf R-r}_i\right| }=Z_Pe^2%
{1/R\;\;\text{if\ \  }r_i<R \atopwithdelims\{\} 1/r_i\;\;\text{if\ \ \ }r_i\geq R}
\;\;\;\text{(monopole approx.)\ ,} \\
&=&\sqrt{\frac{4\pi }3}Z_Pe^2Y_{10}(\widehat{{\bf r}})%
{r_i/R^2\;\;\text{if\  }r_i<R \atopwithdelims\{\} R/r_i^2\;\;\;\text{if\ \ \ }r_i\geq R}
\;\;\;\text{(dipole approx.)\ ,}
\end{eqnarray}
The first equation is valid in the monopole approximation and the second
equation is valid in the dipole approximation, in a head-on collision. $r_i$
are the positions of the atomic electrons, and $R$ is the distance between
the atomic nuclei.

Using hydrogenic wavefunctions and considering only the atomic polarization
arising from the transitions from the ground state, $\Phi _{nlm}\equiv \Phi
_{100}$, and the $s$-state, $\Phi _{nlm}\equiv \Phi _{200}$, we get

\begin{equation}
\left\langle \Phi _{100}\left| V_C^{mon}({\bf r,R})\right| \Phi
_{200}\right\rangle =\frac{4\sqrt{2}}{27}\frac{Z_TZ_P\ e^2}{a_0}\;f(\chi )\;,
\end{equation}
where $\chi =3Z_TR/2a_0$ and

\begin{equation}
f(\chi )=\left( 1+\chi \right) \exp (-\chi )\;.
\end{equation}
For the dipole excitations, considering transitions from the ground state, $%
\Phi _{nlm}\equiv \Phi _{100}$, and the $p$-state, $\Phi _{nlm}\equiv \Phi
_{210}$, we get

\begin{equation}
\left\langle \Phi _{100}\left| V_C^{dip}({\bf r,R})\right| \Phi
_{210}\right\rangle =\frac 8{9\sqrt{2}}\frac{Z_TZ_P\ e^2}{a_0}\;g(\chi )\;,
\end{equation}
where

\begin{equation}
g(\chi )=\frac 1{\chi ^2}\left[ 8-\left( 8+8\chi +4\chi ^2+\chi ^3\right)
\exp (-\chi )\right] \;.
\end{equation}

For an estimate of the effect of atomic polarizability, we will assume that
the contribution of virtual excitations of the $s$- and $p$-orbit are the
most relevant. For a hydrogen atom $E_2-E_1=10.2$ eV. Using this value in
eqs. (2.21-2.24), also for heavier atoms, we find the corrections due to
atomic polarizability presented in the two last rows of Table I. We see that
atomic polarizability is more important for monopole excitations. The
reasoning here is the same as in the use of the adiabatic approximation for
the effect of electron screening: the monopole field of the combined $%
(Z_P+Z_T)$ atom is stronger than the monopole field of the $Z_P$ atom. Thus,
the contribution of the monopole term dominates over other multipolarities.
However, its effect on the fusion cross section is still small. This agrees
with the hypotheses used in the dynamical calculations \cite{Sho93} of
atomic screening that effects due to atomic excitations, and particularly
for virtual excitations, are small and can be neglected.

\subsection{Conclusions}

In conclusion, we have shown that the vacuum polarization, relativistic
corrections, Bremsstrahlung, and atomic polarization contributions to the
astrophysical S-factor never exceed a few percent, but may be significant in
extrapolating the measured S-factor to lower energies. Although these
contributions are not comparable to that of sub-threshold resonances and
electron screening, they represent some of the many factors that may
contribute to the weak energy dependence of the S-factor. Nuclear
polarization effects, were not included, since they are much smaller than
effects due to atomic polarization, for light targets.

Vacuum polarization effects are the most important from all small
contributions, and sensitive to the inner turning point of the potential
barrier, hence to the diffuseness of the nuclear potential employed.
Although the energy needed to create a virtual $e^{+}e^{-}$ -pair is much
larger than atomic excitation energies, the magnitude of its effect is
greatly compensated by its large matrix element (due to the large overlap of
the electron and positron wavefunctions), contrary to the atomic
polarization cases. For the same reason, nuclear polarization and excitation
should be neglected.

We have shown in this work that none of the corrections beyond the effect of
atomic screening can explain the missing enhancement of the fusion cross
sections in atomic target experiments. As suggested in ref. \cite{Ban96},
this effect might well be due to a wrong assumption on the dependence of the
stopping power on the beam energy.

\bigskip\bigskip
\noindent{\bf Acknowledgments}

\medskip

We thank K. Langanke for useful comments. This work was supported in part by
MCT/FINEP/CNPQ(PRONEX) under contract No. 41.96.0886.00, in part by the U.S.
National Science Foundation Grant No. PHY-9605140, in part by the FAPESP
under contract number 96/1381-0, and in part by the University of Wisconsin
Research Committee with funds granted by the Wisconsin Alumni Research
Foundation.

\newpage
\bigskip
\noindent
\begin{tabular}{|l|l|l|l|l|l|l|l|l|l|}
\hline
Reaction & $E_{\min }$ & $U_e$ & $R_n$ & $1-R_{Scr}$ & $1-R_{VPol}$ & $%
1-R_{rel}$ & $1-R_{Br}$ & $1-R_{at}^{dip}$ & $1-R_{at}^{mon}$ \\ 
& $[keV]$ & $[eV]$ & $[fm]$ &  & $\left[ \times \;10^{-2}\right] $ & $\left[
\times \;10^{-3}\right] $ & $\left[ \times \;10^{-3}\right] $ & $\left[
\times \;10^{-5}\right] $ & $\left[ \times \;10^{-1}\right] $ \\ \hline
$D(d,p)T$ & 1.62 & 20 & 4.3 & $0.164$ & $-0.95$ & 0.17 & 0.54 & 1.01 & 0.246
\\ \hline
$^3He(d,p)^4He$ & 5.88 & 119 & 4.3 & $0.331$ & $-1.60$ & 0.47 & 1.12 & 0.39
& 0.314 \\ \hline
$D(^3He,p)^4He$ & 5.38 & 113 & 4.3 & $0.364$ & $-1.58$ & 0.47 & 1.00 & 2.11
& 0.357 \\ \hline
$^3He(^3He,2p)^4He$ & 25 & 292 & 3.0 & $0.196$ & $-3.14$ & 1.75 & 0.58 & 0.35
& 0.321 \\ \hline
$^6Li(p,\alpha )^3He$ & 10.74 & 186 & 3.0 & $0.258$ & $-1.82$ & 1.07 & 1.36
& 0.30 & 0.360 \\ \hline
$^7Li(p,\alpha )^4He$ & 12.70 & 186 & 4.3 & $0.198$ & $-1.88$ & 1.04 & 1.28
& 0.17 & 0.284 \\ \hline
$^6Li(d,\alpha )^4He$ & 14.31 & 186 & 3.0 & $0.218$ & $-2.32$ & 0.72 & 0.71
& 0.15 & 0.313 \\ \hline
$H(^6Li,\alpha )^3He$ & 10.94 & 186 & 3.0 & $0.250$ & $-1.82$ & 1.07 & 1.23
& 2.55 & 0.350 \\ \hline
$H(^7Li,\alpha )^4He$ & 12.97 & 186 & 4.3 & $0.191$ & $-1.88$ & 1.04 & 1.17
& 1.42 & 0.275 \\ \hline
$D(^6Li,\alpha )^4He$ & 15.89 & 186 & 3.3 & $0.184$ & $-2.34$ & 0.71 & 0.35
& 0.91 & 0.262 \\ \hline
$^{10}B(p,\alpha )^7Be$ & 18.70 & 346 & 3.3 & $0.376$ & $-2.38$ & 2.03 & 1.45
& 0.57 & 0.758 \\ \hline
$^{11}B(p,\alpha )^8Be$ & 16.70 & 346 & 2.0 & $0.462$ & $-2.36$ & 2.00 & 2.13
& 0.85 & 0.906 \\ \hline
\end{tabular}

\bigskip

{\bf Table Caption:}

Lowest experimental energies, $E_{\min }$, energy corrections \cite{Bra90}
due to the screening by the atomic electrons, $U_e$, nuclear radii, and
correction factors for the nuclear reaction: (a) due to atomic screening, $%
1-R_{Scr}$, (b) vacuum polarization, $1-R_{VPol}$, (c) relativity, $1-R_{rel}
$, (d) Bremsstrahlung, $1-R_{Br},$ and (e) atomic polarization, $1-R_{at}.$

\bigskip 
{\bf Figure Caption}\\

{\bf Fig. 1} - Comparison between the Coulomb potential and the vacuum
polarization potential as a function of the nuclear separation distance for $%
Z_1Z_2=1.$ The vacuum polarization potential has been multiplied by a factor
1000 in order to be visible in the same plot.

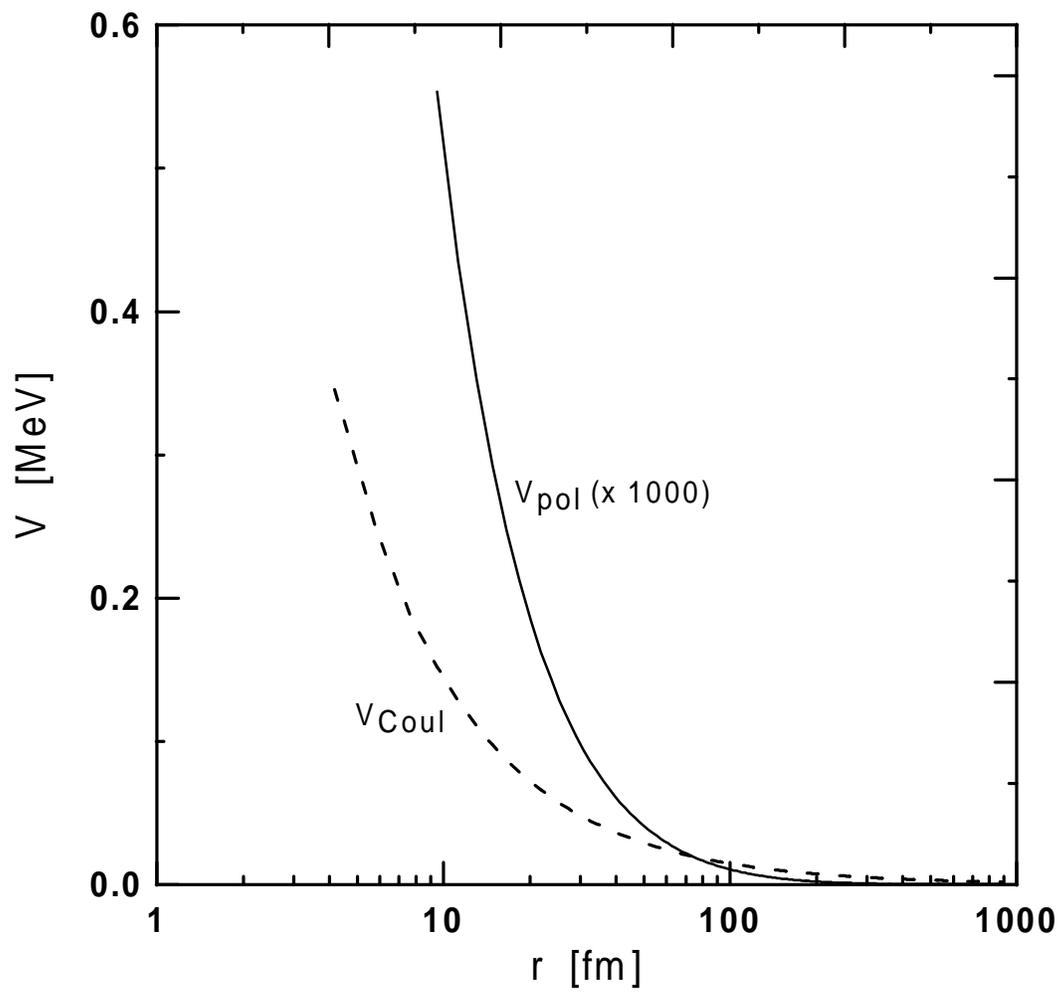
\end{document}